\pdfoutput=1
\documentclass[letterpaper,twocolumn,10pt]{article}
\usepackage{usenix2019_v3}

\usepackage{epsfig}
\usepackage{xspace}
\usepackage{xcolor}
\usepackage{amsthm}
\usepackage{amsmath}
\usepackage{amssymb}
\usepackage{mathrsfs}
\usepackage[all]{xy}
\usepackage{graphicx}
\usepackage{float}
\usepackage[caption = false]{subfig}
\usepackage{times}

\newcommand{\msg}{\ensuremath \mathsf{t}\xspace}

\newcommand{\seqnume}{\ensuremath \mathsf{s_e}}

\newcommand{\id}{\ensuremath \mathsf{id}}

\newcommand{\DE}[1]{\ensuremath \mathsf{REC_{#1}}}
\newcommand{\de}[1]{\ensuremath \mathsf{rec_{#1}}}

\newcommand{\seqnum}{\ensuremath \mathsf{s}}

\newcommand{\naive}{\emph{naive}\xspace}

\newcommand{\putmsg}{\code{putmsg}\xspace}
\newcommand{\getmsg}{\code{getmsg}\xspace}
\newcommand{\mysection}[1]{\vspace{-.15cm}\section{#1}\vspace{-.1cm}}
\newcommand{\mysubsection}[1]{\vspace{-.15cm}\subsection{#1}\vspace{-.1cm}}

\newcommand{\myparagraph}[1]{\vspace{-0.5em}\paragraph{#1}}
\newcommand{\eg}{\hbox{\emph{e.g.},}\xspace}
\newcommand{\ie}{\hbox{\emph{i.e.},}\xspace}

\usepackage{listings}
\usepackage{algorithm}
\usepackage{algpseudocode}

\newcommand{\TOOL}[0]{\textsc{Fidelius}\xspace}
\newcommand{\mycomment}[1]{}
\newcommand{\code}[1]{\textsc{#1}}
\newcommand{\javacode}[1]{\texttt{#1}}
\newcommand{\tuple}[1]{\ensuremath \langle #1 \rangle}

\algnewcommand{\algorithmicgoto}{\textbf{go to}}%
\algnewcommand{\Goto}[1]{\algorithmicgoto~\ref{#1}}%

\newcommand{\squishcount}{
   \begin{list}{\arabic{enumi})}
     { \usecounter{enumi}
       \setlength{\itemindent}{.5em}
       \setlength{\parskip}{0pt}
      \setlength{\itemsep}{0pt}      \setlength{\parsep}{1pt}
      \setlength{\topsep}{1pt}       \setlength{\partopsep}{0pt}
      \setlength{\leftmargin}{0.7em} \setlength{\labelwidth}{1em}
      \setlength{\labelsep}{0.5em} } }

\newcommand{\countend}{
  \end{list}}
\newcommand{\squishlist}{
   \begin{list}{$\bullet$}
     { \setlength{\itemindent}{.3em}       \setlength{\parskip}{0pt}
       \setlength{\itemsep}{0pt}      \setlength{\parsep}{1pt}
      \setlength{\topsep}{1pt}       \setlength{\partopsep}{0pt}
      \setlength{\leftmargin}{0.7em} \setlength{\labelwidth}{1em}
      \setlength{\labelsep}{0.5em} } }
\newcommand{\squishend}{
    \end{list}  }

\usepackage{url}
\usepackage{hyperref}

\begin{document}



\date{}

\title{\Large \bf Securing Smart Home Devices against Compromised Cloud Servers}

\author{
{\rm Rahmadi Trimananda, Ali Younis, Thomas Kwa, Brian Demsky}\\
University of California, Irvine
\and
{\rm Harry Xu}\\
UCLA
} 

\maketitle
\begin{abstract}
This paper presents \TOOL---a runtime system for secure cloud-based
storage and communication even in the presence of compromised
servers. \TOOL's design is tailored for smart home systems that
have \emph{intermittent Internet access}. In particular, it supports
local control of smart home devices in the event that communication
with the cloud is lost, and provides a consistency model using
transactions to mitigate inconsistencies that can arise due to network
partitions. We have implemented \TOOL, developed a smart home
benchmark that uses \TOOL, and measured \TOOL's performance and power
consumption.  Our experiments show that compared to the commercial Particle.io framework, \TOOL reduces more than \textbf{50\%} of the data communication time and increases battery life by \textbf{2$\times$}.
Compared to PyORAM, an alternative (ORAM-based) oblivious storage implementation, \TOOL
has \textbf{4-7$\times$}
faster access times with \textbf{25-43$\times$} less data transferred. 

\end{abstract}

\mysection{Introduction}\label{sec:intro}
\mycomment{
Smart home devices have received increasing attention due to their ability to
provide smart functionality include lights,
door openers, water flow sensors, smart meters, IP-enabled cameras, motion sensors, and
door locks. These devices can be easily attacked, leading to privacy leaks or even physical home damages~\cite{fernandes-oakland16}. Much work~\cite{vigilia, flowfence,he-sec18} has been done at various layers of the stack to protect such client-side applications and devices.
}

In contrast to extensive research on client-side smart home security, the security of the cloud servers
that control these client-side devices has received less attention from researchers.
This is particularly concerning---e.g., recent work shows that if an attacker can control enough high-wattage
IoT devices, the attacker can cause power grid
failures~\cite{powergrid1}.
Implementing attacks \emph{at scale} by compromising individual smart home devices is not
straightforward due to the sheer number of devices that must be hacked.  
Thus, compromising cloud servers can be a more practical
approach.  In fact, the year of 2018, alone, has seen a huge number of compromises to cloud servers, including those
operated by many well-known companies (\eg Facebook~\cite{facebook}, Marriott~\cite{marriott}, Sony~\cite{sony}, etc.)

\mycomment{
Such as
Facebook~\cite{facebook}, Marriott~\cite{marriott}, Sony~\cite{sony},
Panera~\cite{panera}, Ticketfly~\cite{ticketfly}, Quora~\cite{quora},
Newegg~\cite{newegg}, Rail Europe~\cite{raileurope}, British Airways~\cite{britishairways},
and Target~\cite{target}.
}

In addition to large-scale attacks on
physical infrastructures, there are other concerns with trusting the 
cloud: 
(1) \textit{Privacy:} Smart home devices collect a great deal of
information about  users~\cite{smarthomecollectinginfo1}.
Private data can be leaked if a cloud server is compromised.
(2) \textit{Traffic Analysis:} 
It may be possible to learn information by
observing traffic patterns~\cite{princeton-spying-castle}.
For example, such an analysis could reveal that when the thermostat's
mode transitions (\eg from \textit{Home} to \textit{Away}), it always sends
a packet of a specific length~\cite{copos2016anybody}. 
By watching traffic, attackers could discover whether people are home.

\begin{figure}[!htb]
\vspace{-.5em}
\begin{center}
\includegraphics[scale=0.49]{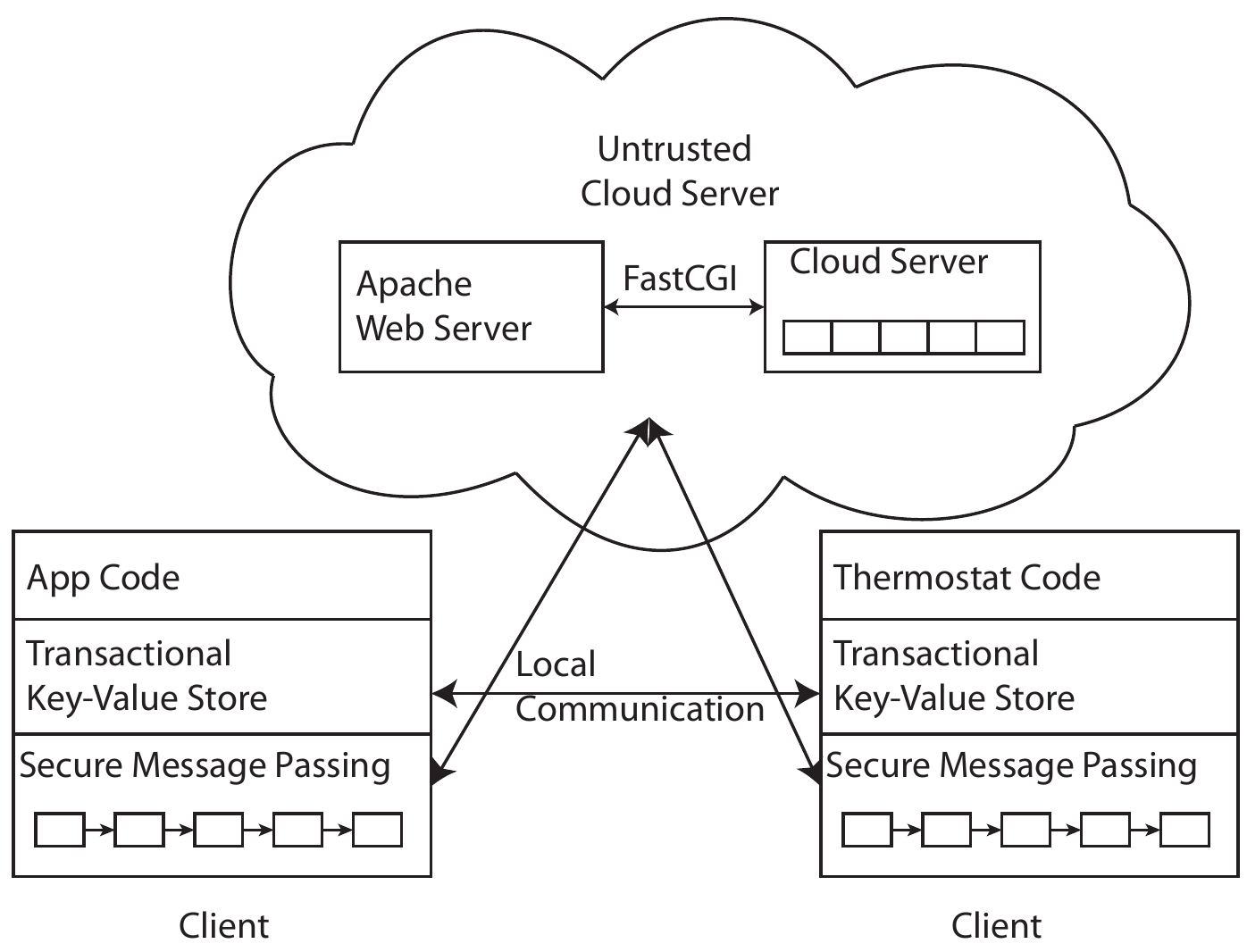}
\end{center}
\vspace{-1em}
\caption{System Overview\label{fig:overview}}
\vspace{-2em}
\end{figure}

\myparagraph{Our Approach: \TOOL}
Motivated by these security and privacy concerns, this paper
presents \TOOL.
\TOOL provides consistency and security
to smart home devices even in the presence of compromised servers in realistic smart home environments that include 
the presence of intermittent network accesses.
Figure~\ref{fig:overview} presents an overview of the \TOOL system.
A \TOOL deployment consists of (1) an untrusted cloud-based server
that provides connectivity between clients, (2) any number of clients
that are smart home devices, and (3) any number of clients that are
smartphone apps.  Our server implementation is architected as a
FastCGI server that communicates with the Apache Web Server. 
We focus our work on providing a
secure \emph{key-value storage system} targeted at IoT applications.\footnote{
The Nest thermostat API
communicates data in JSON as a tree of key-value pairs that can be
easily stored using a key-value abstraction.  Similarly, Apple HomeKit
associates a set of properties and corresponding values with each
device.  
}
While existing systems use adhoc protocols for communicating data, the standard
\emph{key-value abstraction} is powerful enough to subsume a
wide-range of adhoc protocols.
We do not impose special 
requirements on server hardware; instead, the \TOOL 
enforcement runs on \emph{each client device}. Clients communicate with the server; in the absence of Internet connectivity, they can also 
communicate with each other locally
to maintain functionality.\footnote{This is already supported in many smart home systems today such as LiFX and Hue light bulbs, WeMO outlets, and TP-Link outlets.}
This design is well matched for the smart-home environment where 
devices are mutually connected in a local home network. 

\myparagraph{Contributions}
This paper makes the following contributions:
\squishlist
\item {\bf Secure Transactional Key-Value Store:}
  It presents a key-value store that provides strong
  security and privacy guarantees even if the server is malicious.
\item {\bf Local Control:} It presents an algorithm supporting local control of smart home devices
when  connectivity is lost.
\item {\bf Transactional Programming Model:}  It presents developers with a transactional model to abstract consistency and availability tradeoffs that arise from network partitions.
\item {\bf Evaluation:} 
Compared to Particle.io, \TOOL reduces more than \textbf{50\%} of the data communication time and increases battery lifetime by \textbf{2$\times$}. Compared to PyORAM,  \TOOL has \textbf{4-7$\times$} faster access times with \textbf{25-43$\times$} less data transferred. 
\squishend


\mysection{Threat Model}\label{sec:threat}

\TOOL focuses on preventing attackers from using compromised servers to mount large scale attacks on smart home devices.
The problem of
client security for smart home devices is thus outside of the scope of this
paper.\footnote{ \TOOL can be easily used together with any existing client-side security enforcement technique to provide a \mycomment{complete} comprehensive
solution to the problem of smart-home IoT security.}
We assume that the adversary has leveraged other vulnerabilities to
obtain complete control over the cloud server that is used by the
devices to communicate.  We assume that the adversary has knowledge
of \TOOL but does not know the secret key to which only the clients have access.

\TOOL provides the following guarantees, which are formalized in the technical report~\cite{technicalreport}: 
\squishcount
\item \TOOL provides oblivious privacy.  
\TOOL does not leak any information from data access
patterns.  However, \TOOL may leak information from 
timing.

\item \TOOL guarantees the integrity of the message passing
layer.  It ensures that any violation of integrity is detected by some
client and that even in the case of a detected integrity violation,
the consequences for a client is limited to the same set of
behaviors that are allowed by the non-determinism from the network, \ie
could have arisen with a honest server receiving messages from clients in a different order.
\countend

A malicious server can implement a denial-of-service (DoS)
attack by not forwarding messages.  A malicious server can also
partition clients into disjoint groups that it does not permit to
communicate.  This attack is impossible for a client to distinguish
from a network failure that prevents the server from
communicating with a subset of clients, and thus \TOOL provides
fork consistency~\cite{sundr} by default.  
\mycomment{
Section~\ref{sec:consistency} discusses how \TOOL can
provide strong consistency guarantees than fork consistency.}


\mysection{The \TOOL System}
\label{sec:example}

A key challenge in the design of \TOOL is how to handle the wide-range of
possible malicious behaviors from the server  while at the same time
support local control during Internet outages as well as intermittent availability of power-constrained devices.
To separate the concerns of 
handling malicious behaviors and implementing the key-value-store functionalities,
we architected \TOOL as two layers:
(1) a message passing layer that guarantees
consistent message delivery in the presence of  malicious
servers.  This layer handles all issues related
to malicious behaviors and does not implement any key-value-store functionalities; and
(2) a transactional key-value store built on top of the
message passing layer.
The app code runs on top of these two layers in each client---app developers can use the \TOOL API library in their code. On smart home devices, this code is the device controller code.  On smartphones, this is the actual app code. 

\mycomment{
Recall that \TOOL is architected as a key-value store built on top of a message passing layer.  We begin our presentation of \TOOL with the 
message passing layer. The key-value store layer (presented in Section~\ref{sec:kvstore}) uses the message passing layer to send data between clients via the cloud server.  
Data from the key-value store is stored in records that are part of the messages.   
Conceptually, the functionality of the message passing algorithm is simple: the server algorithm maintains a queue of all messages that are currently in use; 
the client algorithm supports (1) sending messages to the server
queue and (2) requesting a copy of the new messages in the current queue.  In addition to forwarding messages between clients, the server stores enough state so that new or long-absent clients can completely reconstruct the key-value store using the state in the server's queue.

The complexity in the algorithm arises in ensuring that compromised or otherwise malicious
servers do not violate the following correctness properties:
(1)  all clients receive a consistent set of messages, (2)  the server delivers all messages that it acknowledges, and
(3)  the server cannot forge messages or replay old messages.  The client algorithm detects and
signals an error if the server violates these properties.
}

\mysubsection{Message Passing Layer}\label{sec:message-passing}
Conceptually, the functionality of the message passing algorithm is simple: the server algorithm maintains a queue of all messages that are currently in use; 
the client algorithm supports (1) sending messages to the server
queue and (2) requesting a copy of the new messages in the current queue.  In addition to forwarding messages between clients, the server stores enough state so that new or long-absent clients can completely reconstruct the key-value store using the state in the server's queue.

\begin{figure}[!htb]
\begin{center}
\includegraphics[scale=0.43]{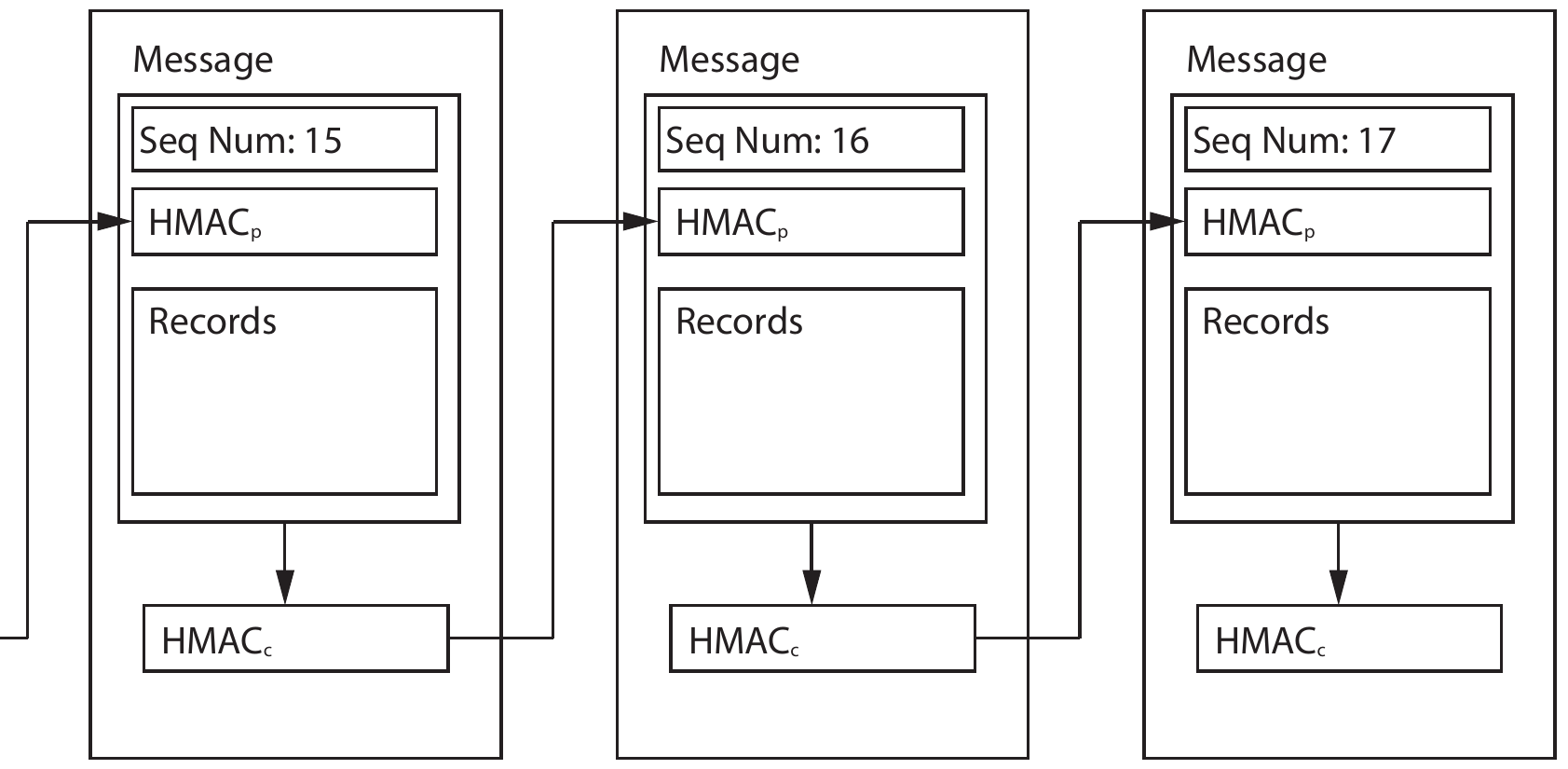}
\end{center}
\vspace{-.5em}
\caption{Message Chain\label{fig:messagechain}}
\vspace{-1em}
\end{figure}

The \TOOL message passing layer uses a message chain to totally order
communications between clients that are sent via the cloud server.  Figure~\ref{fig:messagechain}
presents the structure of the message chain.  Each message in the
chain has a fixed size and contains (1) a globally unique sequence number, which is used to reference and order messages; (2) a
keyed-hash message authentication code (HMAC) of the previous message
in the chain, which is used to ensure the integrity of chain; and (3) a set of records, which can contain either data from the key-value store layer or control information that is needed to thwart attacks---we insert as much data as possible until the entire message reaches a certain fixed size specified through the \TOOL API.  Finally, each message contains an
HMAC of itself, which ensures that the message cannot be modified.
The HMACs are generated using a key known only to the clients to
ensure that the server cannot forge messages.  When a message is
transported between clients, it is always encrypted using the
transport key.  Together, this ensures that the server cannot know the
contents of messages nor can it modify 
them.\footnote{
Although the entire process might seem heavy for IoT devices that have limited resources, our evaluation energy consumption suggests that \TOOL can still perform efficiently (see Section~\ref{sec:eval}).}

\mycomment{
A \naive message chain would eventually consume all memory on the
devices and server.  Thus, \TOOL is designed such that the last $n$
messages in the chain have enough information to reconstruct the state
of the key-value store.  This enables the server to truncate the message
chain.  An important secondary benefit of truncating the message chain
is that the amount of work a client must perform when it 
wakes up after a long sleep is bounded by $n$.

Records in the key-value store can potentially still be in use (or \emph{live}) once they reach the end of the chain.  Thus, a
consequence of our design choice is that \TOOL must reinsert (or refresh) such records into the message chain before they are
truncated.  Section~\ref{sec:mpsend} provides details about how \TOOL determines whether a record is still live.
}

\myparagraph{Server Algorithm}
\mycomment{
The server algorithm implements
a shared, fixed-size message queue where clients can (1) insert new
messages and (2) read the contents of existing messages.  It is
important to note that the server is not trusted to correctly execute
this algorithm---clients will detect if the server incorrectly
executes the server algorithm.  However, if the server does not
correctly execute the server-side algorithm, it can prevent
transactions from committing.
}
The server maintains a queue of the last $n$ messages it has received.  Each message has a fixed-size encrypted block for which the server does not have the decryption key and a plain text sequence number $\seqnum$.
The server algorithm supports two requests: 
(1) \putmsg: add a new message\footnote{The server could first authenticate participating clients to avoid amplification attacks. However, this is out of the scope of this paper.} and 
(2) \getmsg: request messages that are currently
in the queue.  
Before adding a new message, it checks that its sequence number $\seqnum$ is one
greater than the previous sequence number.  If not, it sends a
rejection message back to the client that contains all messages in the
queue with a sequence number that is greater than or equal to
$\seqnum$.   Note that the clients must verify that the server correctly assigns these sequence numbers as we do not trust the server.
If a client detects that the key-value store state does not fit in the current queue, it can request that the server changes the size of the queue.
If a partial message is received, e.g., due to a network failure, the
server ignores the partial message.  If 
queue is full when a new message arrives, the server drops the oldest
message.  The client algorithm is responsible for ensuring that the information in the oldest message has been moved to a newer message before the message is dropped. 
A request for messages in the queue contains a sequence number $\seqnum$ that is one larger than the most recent message the client has received.  When the
server receives this request, it sends all messages
in its queue with sequence numbers that are equal or larger than
$\seqnum$ such that the client now has a up-to-date copy of the message chain.

\myparagraph{Client Algorithm}
\mycomment{$\msg = \mathsf{\tuple{s, E(Dat_s)}}$}
Clients communicates by sending and receiving messages through the server.
\mycomment{
A message $\msg$ consists of the sequence number $s$ and the encrypted
message body $\mathsf{E(Dat_s)}$.  The message body $\mathsf{Dat_s}$ consists of the sequence number $\seqnume$, the machine
ID $\id$ of the sender, the HMAC $\mathsf{hmac_p}$ from the previous
message, the set $\DE{}$ of records that contain the contents of the
message, and the HMAC $\mathsf{hmac_c}$ of the current message.
Messages have two copies of the sequence number as the server needs
access to the sequence number and the clients need a copy that they
can authenticate.  }
Our implementation uses the CTR mode of AES to encrypt messages.  A unique counter value
is sent to the server along with the message.  Clients share two
secrets that allow them to authenticate and encrypt messages: a
message authentic key 
and a message encryption key. These are not known to the server.

\mycomment{
Our presentation of the client algorithm is structured as follows.
Section~\ref{sec:recvalidation} presents the
validation checks that clients perform when they receive messages.
Section~\ref{sec:mpsend} presents
the algorithm clients use to send messages.
}

When the message passing layer receives new messages from the server
(or they are created locally), the client decrypts and validates the
messages.  Records in messages at the message passing layer are used to both communicate
information for the key-value store layer and to thwart 
attacks that a malicious server may otherwise
perform.  
The proof in the technical report~\cite{technicalreport} shows that these checks suffice to thwart all server attacks: 
(1) \textit{Message Chain Integrity:}
To ensure the integrity of messages, after
decrypting a message the client
compares: computed HMAC against the stored
HMAC, $\seqnum$ from the server against $\seqnum$ in the message, and the HMAC of the current message against the HMAC of the message immediately proceeding
it, if present, in the queue. These are used to ensure that the server has not
manipulated the message chain.
(2) \textit{Detecting Dropped Messages:}
A malicious server could potentially acknowledge a message from a
client and then silently drop the message while a client is offline,
replacing it with a message with the same sequence number from another
client. To ensure that clients always detect
such dropped messages, the client algorithm uses {\it last message
records} to track the last message sent by itself and other clients so
that it can detect whether the server has dropped messages.  A last
message record consists of a
machine ID 
and the sequence number 
of the last
message sent by that machine.  Last message records are only 
inserted if the most recent message from
a given machine ID is about to be evicted from the queue. 
(3) \textit{Detecting Reused Rejected Messages:}
The client algorithm uses {\it rejected message records} to detect if
the server attempts to use a rejected message.  A rejected message
record consists of the machine ID 
$\id$ 
of the
client that sent the rejected messages, the lowest sequence number
$\mathsf{s_{low}}$ 
and highest sequence number 
$\mathsf{s_{high}}$ 
of
the range of rejected messages, and the sequence number
of the first message that contained the rejected
message record.
To prevent this attack, each client verifies that each message in the
queue that has a sequence number that falls within the range of 
$\mathsf{s_{low}}$ and $\mathsf{s_{high}}$ has a machine ID $\id$ that 
is not equal to $\id$. 
The algorithm keeps a rejected message record 
live until all clients have seen it (as implicitly acknowledged by sending 
a newer message).\footnote{Since \TOOL does not keep the entire message chain, 
the server could temporarily fork the message chain and then move clients
from one fork to another if the client is offline for a sufficiently
long time such that all the messages the client has seen have been
evicted from the message chain. In the other fork, the server could
could send dropped or rejected messages.}
(4) \textit{Detecting Missing Messages:}
A malicious server could also attempt to send fewer messages than the
queue currently holds by omitting the oldest messages (noting that the 
message chain integrity checks prevent dropping messages in the middle).  
This would cause clients to compute the
key-value store state using an incomplete set of messages.
To prevent this, \TOOL uses {\it queue size record} to
store the current maximum capacity of the message queue and enables
clients to compute how many messages they should have received 
to validate that the correct number of messages was received.  

The message passing layer sends a message and performs the following bookkeeping tasks:
(1) \textit{Determine Which Records Must be Refreshed:}
If a record
$\de{}$ is live (the system is still using the information
they contain) then it must be refreshed by reinserting the record
into the queue before the message with that record leaves the queue.
A record $\de{}$ is dead iff one of the following conditions is true:
(i) a refreshed version of $\de{}$ is present in the 
queue,\footnote{A resize check is performed first to check if the number of messages
with records which are live in the queue exceeds a randomized resize threshold.
If so, a new queue size is calculated and a new queue size record
with the new size is inserted into the message.}
(ii) $\de{}$ is a queue size record and there is a newer queue size record,
(iii) item $\de{}$ is a rejected message record and all clients
have seen it as proven by having inserted a message into 
the queue with a $\seqnum$ that is greater than the $\seqnum$ of $\de{}$, or
(iv) $\de{}$ is a last message record and a client with the same $\id$ has inserted a 
newer message into the queue.
(2) \textit{Construct the Record Section of the Message:}
Constructing the record section of a message involves (i) checking if the queue needs
resizing, (ii) generating and inserting a rejected message record if needed,
(iii) refreshing older live records before they are evicted,
(iv) inserting the record from the key value store layer, and
(v) filling unused space in the 
record section with old live records in need of
a refresh.
(3) \textit{Send the Message:} The client next sends the message
to the server.  If the server rejects the message because it has an
old sequence number, then the server returns the newer messages to
prove that the rejected message was in fact old.  The client then
processes all of the new messages that were returned by the server and
will generate a rejected message record to communicate the fact that
it sent a message that was rejected.  If the server accepts the
message, then the client's local state is updated using the contents
of the newly created message.  If sending the message fails due to
network issues then the message passing layer has to defer its
determination of whether other clients received the message and the
transactional key-value store layer is informed that a network failure
occurred.

\myparagraph{Information Leakage}
A reader may note that the pattern of \putmsg and \getmsg requests
can leak access pattern information.  This issue
can be addressed in several different ways.  We modified
the update procedure to always follow a \getmsg call with a \putmsg call (simply sending a message that refreshes entries if there is no request to send).
Nevertheless, \TOOL does not prevent leaks
via timing channels.  For some settings, it is feasible to
configure \TOOL to not leak information by sending data on a
fixed schedule.\footnote{
We have observed that many smart home devices frequently send messages
to the cloud even when idle: the Nest thermostat sends a packet every
5 seconds, the LiFX light bulb every 7 seconds, and the D-Link smart
plug every 1.5 seconds.}

\myparagraph{Strengthening Consistency Guarantees}\label{sec:consistency}
Although we believe that fork consistency is sufficient for smart home systems,  
\TOOL supports extensions to counter forking attacks and achieve strong consistency~\cite{technicalreport}.

\mysubsection{Transactional Key-Value Store}\label{sec:kvstore}
\TOOL's transactional key-value store is built on top of the
message passing layer.
This layer
(1) arbitrates/commits transactions, and
(2) updates/reads key-value pairs.

\mycomment{
\begin{figure}[!htb]
\vspace{-1.1em}
\includegraphics[scale=0.35]{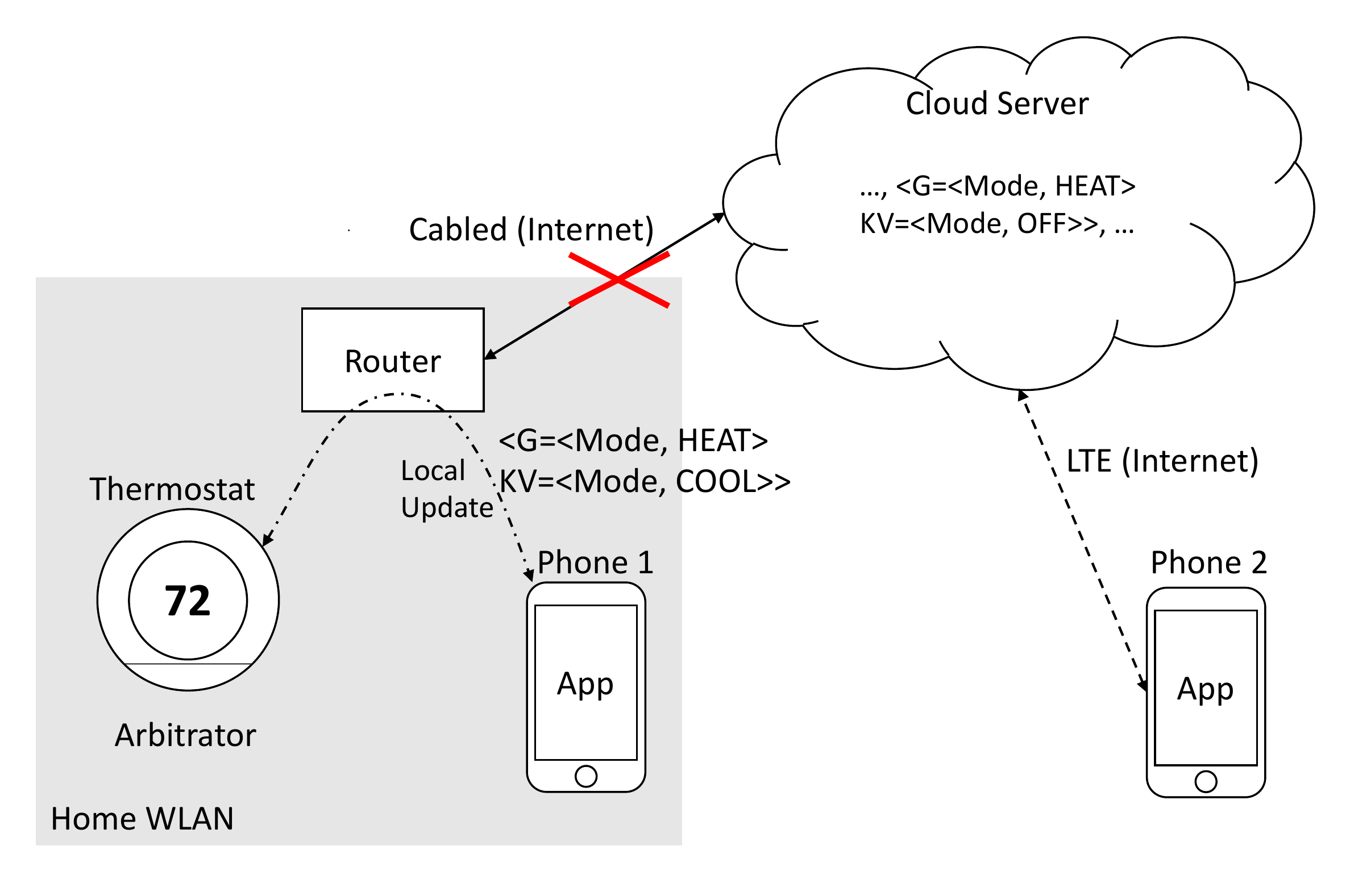}
\vspace{-1.5em}
\caption{\label{fig:conflict}Example Arbitration Scenario}
\vspace{-0.5em}
\end{figure}
}

\myparagraph{Arbitrating/Committing Transactions}
Intermittent home Internet connectivity complicates committing transactions.  There is the potential for concurrent local updates and remote updates to conflict.  In the case of an Internet outage, the cloud server would not even be aware of local updates. 
For example, a remote smartphone attempts to change the thermostat mode from
HEAT to OFF with a transaction.  At the same time, a local smartphone tries to change the mode from HEAT to COOL with a transaction. These two transactions conflict and only one transaction can commit and thus we need an \emph{arbitrator} to decide which one commits.  
To allow local control during an Internet outage, the arbitrator of this transaction must be local to the thermostat and the obvious choice is the thermostat itself.

\myparagraph{Updating/Reading Key-Value Pairs}
\TOOL updates key-value pairs through a \code{Put} function. 
These updates are stored locally until the transaction is committed.
\TOOL supports a relaxed transactional model for reads.
In many cases it is not
important that a transaction reads the absolutely latest value from a
sensor. 

\mycomment{
\TOOL provides two get functions for this purpose:
(1) \code{GetCommittedAtomic:} This accessor function ensures
serializability (\ie isolation).
(2) \code{GetCommitted:} 
This function allows accesses to violate transaction
serializability without causing a transaction to abort.  This can be
useful to read data values for which the application only needs
approximately current information.
}

\mycomment{
In many cases, it is desirable to allow interactive updating of a
device's state without waiting for the device to contact the cloud.
Even if the network connectivity is lost, such transactions are unlikely 
to actually conflict.  \TOOL supports a speculative 
get operation that reads values from pending transactions.
}

\mysection{Evaluation}\label{sec:eval}

We developed both C++ and Java implementations of the \TOOL client and
a C++ implementation of the \TOOL 
server.\footnote{
Most existing smart home systems are closed source
and it is not clear what guarantees are provided when there are conflicts.
Thus, we cannot implement our approach to securing smart home devices against malicious cloud servers directly on these devices. 
Instead, we implemented a complete system (with an API library) that uses transactions to support local communication with clear consistency properties.
}
The server for all experiments was a 3.5GHz Intel Xeon E3-1246 v3 with
32GB of RAM.  We have evaluated \TOOL by (1) developing a test bed
system, (2) comparing with the commercial Particle
cloud in terms of energy usage and privacy, and (3) comparing with a Path ORAM implementation~\cite{stefanov-ccs13b}.


\myparagraph{Test Bed}
Our test bed system simulates a medium scale smart home deployment
with 16 devices.  We have two classes of smart home devices in our
test bed: (1) 15 low-power nodes that use hardware similar to
smart home devices that run on batteries for many months (8 Particle
Photons with temperature and humidity sensors, 4 Particle Photons with
magnetic door sensors, 3 Particle Photons with IR-based motion
sensors) and (2) one mid-range node that uses hardware that is more
typical of smart home devices that run on wall power 
(a Raspberry Pi 1 that controls 2 LiFX smart light bulbs).\footnote{
The Particle Photon is a low-end 
IoT hardware development kit with a 120 MHz ARM Cortex M3
processor, 1MB of flash, and 128KB of RAM that supports 802.11b/g/n
WiFi.}
These specifications are similar to the hardware that appear in commercial devices.  
\mycomment{
For comparison, the Nest Protect battery powered smoke detector
uses a 100 MHz Cortex M4 processor as a primary processor (with 2
other ARM Cortex processors) and thus the Particle Photon is roughly
equivalent to processors in the Nest Protect.
The Raspberry Pi 1 has a 700MHz single core ARM11 processor.  For comparison,
the Nest thermostat contains a 1GHz single core Cortex A8 which is
$\sim2.5\times$ faster than the Raspberry Pi.  Note that we had to use
development kits to run \TOOL since off-the shelf systems are closed source and not modifiable.
}

We quantified the CPU overhead of \TOOL for typical smart home device activity (controlling LiFX light bulbs) on a smart home class CPU (a Raspberry Pi 1) to verify that \TOOL incurs acceptable overheads. 
For this experiment, the LiFX arbitrator had
a 2.3\% CPU utilization and the Android app had below 1\% CPU utilization.  

To validate that \TOOL detects attacks,
we used a malicious server that implements three
different attacks. In the first attack, the server assigns the sequence
number to multiple messages. In the second attack, the server inserts
a rejected message into the queue. In the third attack, the server
swaps message sequence numbers of accepted messages.  \textit{The clients detected all attacks.}

\mycomment{
\begin{table}[th]
  \vspace{-1em}
  \centering
  \begin{center}
  { \footnotesize
  \begin{tabular}{| l | r | r | r | r | r | r |}
    \hline
        Sensor & \multicolumn{3}{| c |}{Particle cloud} & \multicolumn{3}{| c |}{Fidelius cloud}\\
    \cline{2-7}
        Type& Uptime & Energy & Life-  & Uptime & Energy &Life-\\
        & (\textit{s}) &(\textit{mWh}) &time& (\textit{s}) & (\textit{mWh})& time\\
    \hline
        Temp./hum. & 5.55 & 0.79 &1y 4m&  2.46 & 0.35& 2y 2m\\
        IR-based & 5.12 & 0.73 & N/A& 2.19 & 0.31&N/A\\
        Magnetic & 5.10 & 0.73 & N/A& 2.18 & 0.31&N/A\\
    \hline
  \end{tabular}
  }
\end{center}
\caption{Energy consumption of sensors.  The time and energy reported is the time and energy taken to wake up, join the WiFi network, perform a reading, and  report the reading to the cloud. 
For \TOOL, reporting the reading to the cloud involved performing all \TOOL checks and then performing a transaction to send the reading to the \TOOL server (a remote server comparable to the Particle cloud).
Battery lifetimes assume one sensor reading is reported on average per an hour.
\label{tab:energyconsumption}}
\vspace{-2em}
\end{table}

Our Particle Photons
wake up, read the sensor, and then go to deep sleep.
The humidity/temperature sensors are read on average once an hour
while the other sensors wake up the Particle Photon on demand.  
}
\myparagraph{Particle Cloud---Energy and Privacy}
We evaluated \TOOL against the Particle cloud framework as the Particle cloud framework has an easily available SDK that has been used to develop many commercial IoT products.  
We configured a Particle Photon to read a
humidity/temperature sensor and powered it with 3 AA
batteries.  When using the Particle cloud, it
takes at least 5.55 seconds to wake up, connect, publish value, and
go back to sleep.  The total energy consumed for the wake-up time
of 5.55 seconds is 0.79mWh.
Assuming that it wakes up every hour to do one
measurement and goes back to sleep, the total energy consumed is 1.17mWh
each hour for deep sleep plus reporting a measurement.  
Three Energizer AA Lithium batteries provide approximately 13.5Wh and thus,
could support measurements for 1 year and 4 months.  When using 
Fidelius server, it takes only 2.46 seconds to complete the same
measurement cycle. Hence, its hourly power consumption is 0.73mWh,
and it could support measurements up to 2 years 2 months.

\mycomment{
\begin{table}[th]
	\begin{tabular}{cc}
		\begin{minipage}{.5\linewidth}
		\centering
		\footnotesize
		  \begin{tabular}{| c | c |}
			\hline
				\multicolumn{2}{| c |}{\javacode{Particle.publish()}}\\
			\hline
				Argument Length & Payload\\
			\hline
				8 & 34\\
				16 & 34\\
				32 & 50\\
				64 & 82\\
				128 & 146\\
				256 & 210\\
			\hline
		  \end{tabular}
		\end{minipage} &

		\begin{minipage}{.5\linewidth}
		\centering
		\footnotesize
		  \begin{tabular}{| c | c |}
			\hline
				\multicolumn{2}{| c |}{\javacode{Particle.function()}}\\
			\hline
				Argument Length & Payload\\
			\hline
				8 & 34\\
				16 & 50\\
				32 & 66\\
				63 & 98\\
			\hline
		  \end{tabular}
		\end{minipage} 
	\end{tabular}
	\caption{Information leakage of \javacode{Particle.publish()} and \javacode{Particle.function()}.\label{tab:informationleakage}}
\end{table}}

The network trace shows that the Particle cloud 
leaks information about the argument lengths.
\mycomment{Table~\ref{tab:informationleakage} presents the results.}
The \javacode{publish()} method, which allows a device to publish results 
on the Particle cloud in a key-value fashion, sends different packet 
lengths for different argument lengths, e.g., a 16-character argument results in 
a 34-byte payload, a 32-character argument results in a 50-byte payload, etc.
The \javacode{function()} method, which allows devices to publish
C code functions on the Particle cloud, behaves in a similar way.
\mycomment{
e.g., a 16-character argument results in a 50-byte payload, a 32-character
argument results in a 66-byte payload, etc.
Moreover, the Particle cloud has knowledge of all requests and must be trusted.
}
For \TOOL, the network trace only shows a \code{getmsg} is always followed by a
\code{putmsg} request with uniformly-sized data blocks.
We randomized the wake-up time of our sensors that need
to give periodic updates---\ie temperature/humidity sensor.
Since every \TOOL request had the exact same traffic pattern, we were not able to determine the type of \TOOL request from the traffic.

\myparagraph{Path ORAM}
\mycomment{
\begin{table}[th]
  \centering
  \begin{center}
  { \footnotesize
  \begin{tabular}{| r | r | r | r | r | r |}
    \hline
        Blocks & \multicolumn{2}{| c |}{Total Time (\textit{s})} & 
        	\multicolumn{2}{| c |}{Time/Access (\textit{s})} & Factor\\  	
    \cline{2-5}
        & Path ORAM & \TOOL & Path ORAM & \TOOL &\\
    \hline
        63 & 45.67 & 10.59 & 0.72 & 0.17 & 4.3$\times$\\
        127 & 119.25 & 21.97 & 0.94 & 0.17 & 5.4$\times$\\
        255 & 294.03 & 43.85 & 1.15 & 0.17 & 6.7$\times$\\
        511 & 503.27 & 86.68 & 0.98 & 0.17 & 5.8$\times$\\
    \hline
  \end{tabular}
  }
  \end{center}
  \caption{Access times for different numbers of blocks.
  \label{tab:accesstimes}}
\vspace{-2em}
\end{table}

\begin{table}[th]
  \centering
  \begin{center}
  { \footnotesize
  \begin{tabular}{| r | r | r | r |}
    \hline
        Blocks & \multicolumn{2}{| c |}{Total Transmission (\textit{kB})} & Factor\\  	
    \cline{2-3}
        & Path ORAM & \TOOL &\\  	
    \hline
        63 & 4,296.58 & 175.21 & 24.5$\times$\\
        127 & 10,887.89 & 353.83 & 30.8$\times$\\
        255 & 25,796.25 & 710.50 & 36.3$\times$\\
        511 & 61,324.40 & 1,422.88 & 43.1$\times$\\
    \hline
  \end{tabular}
  }
  \end{center}
  \caption{Data transfer for different numbers of blocks.
  \label{tab:datatransmission}}
\vspace{-1em}
\end{table}
}

PyORAM, a Python implementation of the Path ORAM
algorithm~\cite{stefanov-ccs13b}.\footnote{
Path ORAM is one of the few cloud-based
oblivious storage system implementations that are publicly available~\cite{pyoram}.
Path ORAM assumes the weaker honest but curious adversarial model.
}
We tried to run PyORAM's Path ORAM on the Raspberry Pi as a client.
\mycomment{
Unfortunately, it could only support 7 blocks of data---it
crashed when more blocks were added.  Because of this, we used a
laptop with a 
2.5GHz Intel Core i5 and 8GB 
of RAM as a client instead. On this machine, PyORAM can be configured 
to cache the top 2 levels of the tree with a height of 8 and up to
511 blocks of data---we believe that real systems could
easily have this many keys.
}
The results show an increasing trend of time per access
for PyORAM as the tree height and number of blocks grow, while for \TOOL
time per access is always steady. For 255 blocks,
\emph{\TOOL was already 7$\times$ faster than PyORAM}---this  
difference is caused by the nature of PyORAM benchmark that is bandwidth 
limited.
As the tree height and number of blocks grow, the total
amount of transferred data also increases faster
than \TOOL---for 511 blocks, \emph{PyORAM transfers more than
43$\times$ the amount of data that \TOOL transfers.}

\mysection{Related Work}\label{sec:related}

Oblivious RAM techniques, first proposed by Goldreich and
Ostrovsky~\cite{obliviousram}, seek to provide RAM~\cite{wang-ccs15,
  ren-usenixsecurity15, 
  stefanov-ccs13b, 
  stefanov-ndss12, liu-asplos15} or cloud
storage~\cite{stefanov-ccs13, stefanov-oakland13, 
  soled-onetwork15} in which the data access patterns
are independent of the computation and thus no
information can be extracted from the locations accessed.
Many
oblivious storage algorithms~\cite{obliviousram, stefanov-ccs13b} assume that
only one device can access the storage and that adversary is merely
honest but curious.  While some work does support multiple
clients~\cite{multiclient} or stronger adversarial
models~\cite{integrityverification}, these approaches assume that the
clients have limited local storage and thus incur bandwidth and/or latency overheads
for each access.

Much previous work on oblivious storage is
unsuitable for our scenario because: (1) it assumes an
honest, but curious adversary, and (2) it does not support multiple
clients accessing shared state.  Some recent work has removed the
single client limitation~\cite{multiclient} and supports stronger
adversarial models~\cite{integrityverification}.  However, most
oblivious RAM techniques incur extra overheads (e.g., latency or bandwidth)
for 
accesses to storage due to a fundamental limitation as a result of
the assumption that clients only have a small amount of storage.
Moreover, existing systems are not designed to support updates to
persistent state when connectivity has been lost and thus cannot
support local device control.

\mycomment{
There is a large body of work on building network systems that can
make communication unobservable
on untrusted infrastructures. Mix networks~\cite{
  kesdogan-ih98, blond-sigcomm15, blond-sigcomm13} were used by the
 private messaging systems to shuffle messages before
delivering them to recipients.
Mailbox systems
techniques~\cite{hooff-sosp15, angel-osdi16, 
borisov-pets16, lazar-osdi16} provide security for messages
retrieved from mailboxes kept at third-party servers.  These
techniques are orthogonal.
}

\mysection{Conclusion}\label{sec:conclusion}

We presented \TOOL, a new secure key-value store, that is
designed to meet the requirements of smart home systems.
We prove that \TOOL is secure, even
in the presence of a malicious cloud server.  
Our experiments show that \TOOL has
good performance, low power consumption, is resilient to attacks, and efficiently runs on
smart home class hardware.  

\mysection{Discussion}\label{sec:discussion}

\myparagraph{Feedback}
\TOOL is a
privacy-preserving infrastructure with a specific context
(\ie key-value store).
We are seeking feedback about the viability
of \TOOL in the real-world setting with respect to 
the current trends for cloud architecture.
We are also looking into future directions to extend
\TOOL for more complex applications beyond key-value
store model (\eg cameras and video streaming devices).

\myparagraph{Controversial Points}
First, \TOOL distributes key among devices. However, this could
be a bad choice because the entire system will be compromised
if a client gets compromised.
Second, \TOOL tries to leverage encryption in the communication
between IoT devices---this might be too heavy for IoT devices
that typically have limited resources (\eg processing power
and memory). While a lot of devices have implemented
their communication protocols on top of secure channels, \eg
TLS, there are still devices that communicate in clear text
for simplicity.
Third, \TOOL protects the privacy of clients by concealing 
the data from the cloud server. However, the current trend is
that cloud servers typically access clients' data for analytics purposes.

\myparagraph{Likely Discussion}
We hope that this paper could generate discussions
with the following questions (with respect to the
controversial points and the feedback we look for):
\textbf{(1)} How viable is \TOOL to provide 
privacy and security for IoT devices with respect to current trends and practices? What could be the better scheme to distribute keys
if we use encryption (with the assumption that the cloud server could be malicious)? Could we really trust the clients?
\textbf{(2)} How could we develop \TOOL better in
the direction that follows the current trend for cloud
analytics? \TOOL may not
be suitable for all IoT applications. However, it
can be effective for certain scenarios, in which the user may not want
the server to comprehend the data, 
\eg when a company uses a public cloud service to run its applications, it needs to protect its users' sensitive information. 
We realize that when using \TOOL, 
data analytics can still be performed
using another machine (as a \TOOL client).

\myparagraph{Open Issues}
\TOOL provides security through encryption and
packets with the same-length. However, there is 
still privacy leak through timing channels. 
Although, an obvious solution is 
traffic injection~\cite{cai2014cs,cai2014systematic},
this could consume a lot of bandwidth: inefficient communication.
Thus, we wonder if there could be a more efficient
solution to secure the timing channels for IoT
devices.

\myparagraph{Circumstances for Failure}
IoT devices typically have limited processing
power and storage, especially for devices that are
most likely to use the key-value store model.
Thus, if there are too many devices that participate
as a \TOOL client, the message chain can be too
long to fit into the device's local storage. As a result, the entire \TOOL system could fail.

{\footnotesize
\bibliographystyle{abbrv}
\bibliography{paper}

\begin{thebibliography}{10}

\bibitem{technicalreport}
Anonymous.
\newblock Securing smart home edge devices against compromised cloud servers.
\newblock \url{https://bit.ly/technical-report}, 2019.

\bibitem{princeton-spying-castle}
N.~Apthorpe, D.~Reisman, and N.~Feamster.
\newblock A smart home is no castle: Privacy vulnerabilities of encrypted {IoT}
  traffic.
\newblock {\em CoRR}, abs/1705.06805, 2017.

\bibitem{marriott}
T.~Brewster.
\newblock Revealed: Marriott's 500 million hack came after a string of security
  breaches.
\newblock
  \url{https://www.forbes.com/sites/thomasbrewster/2018/12/03/revealed-marriotts-500-million-hack-came-after-a-string-of-security-breaches/},
  December 2018.

\bibitem{copos2016anybody}
B.~Copos, K.~Levitt, M.~Bishop, and J.~Rowe.
\newblock Is anybody home? {Inferring} activity from smart home network
  traffic.
\newblock In {\em Security and Privacy Workshops (SPW), 2016 IEEE}, pages
  245--251. IEEE, 2016.

\bibitem{soled-onetwork15}
D.~Dachman{-}Soled, C.~Liu, C.~Papamanthou, E.~Shi, and U.~Vishkin.
\newblock Oblivious network {RAM}.
\newblock {\em {IACR} Cryptology ePrint Archive}, 2015:73, 2015.

\bibitem{obliviousram}
O.~Goldreich and R.~Ostrovsky.
\newblock Software protection and simulation on oblivious {RAMs}.
\newblock {\em Journal of the ACM}, 43(3):431--473, May 1996.

\bibitem{pyoram}
G.~Hackebeil.
\newblock {PyORAM}.
\newblock \url{https://github.com/ghackebeil/PyORAM}, February 2017.

\bibitem{facebook}
M.~Isaac and S.~Frenkel.
\newblock Facebook security breach exposes accounts of 50 million users.
\newblock
  \url{https://www.nytimes.com/2018/09/28/technology/facebook-hack-data-breach.html},
  September 2018.

\bibitem{lamport-toplas82}
L.~Lamport, R.~Shostak, and M.~Pease.
\newblock The {Byzantine Generals Problem}.
\newblock {\em ACM Trans. Program. Lang. Syst.}, 4(3):382--401, 1982.

\bibitem{sundr}
J.~Li, M.~Krohn, D.~Mazi\`{e}res, and D.~Shasha.
\newblock Secure untrusted data repository ({SUNDR}).
\newblock In {\em Proceedings of the 6th Conference on Symposium on Operating
  Systems Design \& Implementation}, 2004.

\bibitem{liu-asplos15}
C.~Liu, A.~Harris, M.~Maas, M.~W. Hicks, M.~Tiwari, and E.~Shi.
\newblock {GhostRider}: {A} hardware-software system for memory trace oblivious
  computation.
\newblock In {\em Proceedings of the Twentieth International Conference on
  Architectural Support for Programming Languages and Operating Systems}, 2015.

\bibitem{multiclient}
T.~Mayberry, E.-O. Blass, and G.~Noubir.
\newblock Multi-client oblivious ram secure against malicious servers.
\newblock Cryptology ePrint Archive, Report 2015/121, 2015.
\newblock \url{http://eprint.iacr.org/2015/121}.

\bibitem{cloudproof}
R.~A. Popa, J.~R. Lorch, D.~Molnar, H.~J. Wang, and L.~Zhuang.
\newblock Enabling security in cloud storage {SLAs} with {CloudProof}.
\newblock In {\em Proceedings of the 2011 USENIX Conference on USENIX Annual
  Technical Conference}, 2011.

\bibitem{ren-usenixsecurity15}
L.~Ren, C.~W. Fletcher, A.~Kwon, E.~Stefanov, E.~Shi, M.~van Dijk, and
  S.~Devadas.
\newblock Constants count: Practical improvements to oblivious {RAM}.
\newblock In {\em 24th {USENIX} Security Symposium, {USENIX} Security 15,
  Washington, D.C., USA, August 12-14, 2015.}, pages 415--430, 2015.

\bibitem{integrityverification}
L.~Ren, C.~W. Fletcher, X.~Yu, M.~van Dijk, and S.~Devadas.
\newblock Integrity verification for path oblivious-{RAM}.
\newblock In {\em IEEE High Performance Extreme Computing}, 2013.

\bibitem{powergrid1}
S.~Soltan, P.~Mittal, and H.~V. Poor.
\newblock Blackiot: Iot botnet of high wattage devices can disrupt the power
  grid.
\newblock In {\em 27th {USENIX} Security Symposium ({USENIX} Security 18)},
  pages 15--32, Baltimore, MD, 2018. {USENIX} Association.

\bibitem{stefanov-ccs13}
E.~Stefanov and E.~Shi.
\newblock Multi-cloud oblivious storage.
\newblock In {\em 2013 {ACM} {SIGSAC} Conference on Computer and Communications
  Security}, 2013.

\bibitem{stefanov-oakland13}
E.~Stefanov and E.~Shi.
\newblock {ObliviStore}: High performance oblivious cloud storage.
\newblock In {\em 2013 {IEEE} Symposium on Security and Privacy, {SP} 2013,
  Berkeley, CA, USA, May 19-22, 2013}, pages 253--267, 2013.

\bibitem{stefanov-ndss12}
E.~Stefanov, E.~Shi, and D.~X. Song.
\newblock Towards practical oblivious {RAM}.
\newblock In {\em 19th Annual Network and Distributed System Security
  Symposium, {NDSS} 2012, San Diego, California, USA, February 5-8, 2012},
  2012.

\bibitem{stefanov-ccs13b}
E.~Stefanov, M.~van Dijk, E.~Shi, C.~W. Fletcher, L.~Ren, X.~Yu, and
  S.~Devadas.
\newblock Path {ORAM}: An extremely simple oblivious {RAM} protocol.
\newblock In {\em 2013 {ACM} {SIGSAC} Conference on Computer and Communications
  Security}, 2013.

\bibitem{wang-ccs15}
X.~Wang, T.~H. Chan, and E.~Shi.
\newblock Circuit {ORAM}: On tightness of the {Goldreich-Ostrovsky} lower
  bound.
\newblock In {\em Proceedings of the 22nd {ACM} {SIGSAC} Conference on Computer
  and Communications Security, Denver, CO, USA, October 12-6, 2015}, pages
  850--861, 2015.

\bibitem{smarthomecollectinginfo1}
Z.~Whittaker.
\newblock Smart home tech makers don't want to say if the feds come for your
  data.
\newblock
  \url{https://techcrunch.com/2018/10/19/smart-home-devices-hoard-data-government-demands/},
  2018.

\bibitem{sony}
Sony pictures hack.
\newblock \url{https://en.wikipedia.org/wiki/Sony_Pictures_hack}.

\end{thebibliography}
}
\end{document}